\documentstyle[12pt]{article}

\begin{document}

\font\fourteenbf=cmbx10 scaled\magstep2
%
%
\def\ns{\normalsize}
\def\ssize{\normalsize}
\def\l{\left}
\def\r{\right}
\def\setu{U_q\left(H_1,H_2,X^\pm\right)}
\def\super{U_qgl\l(1|1\r)}
\def\xpm{X^\pm}
\def\xp{X^+}
\def\xm{X^-}
\def\hi{H_i}
\def\ho{H_1}
\def\htw{H_2}
\def\ki{K_i}
\def\ko{K_1}
\def\kt{K_2}
\def\kii{K_i^{-1}}
\def\koi{K_1^{-1}}
\def\kti{K_2^{-1}}
\def\g{{\it g}}
\def\dif{q-q^{-1}}
\def\un{\hbox{1\hskip-.25em{\rm l}}}
\def\oti{\otimes}
\def\tri{\triangle}
\def\veps{\varepsilon}
\def\CR{{\cal R}}
\def\fact{e^{i{\pi\over 2}\htw}}
\def\facti{e^{-i{\pi\over 2}\htw}}
\def\cuar{{1\over 4}}
\def\qin{q^{-1}}
\def\CRi{\CR^{-1}}
\def\und#1{{\underline{#1}}}
\def\kin{k^{-1}}
\def\etap{\eta^+}
\newcommand{\nn}{\nonumber}

\rightline{}
\vskip 1.5  true cm

\begin{center}
\Large\bf
Quantum and super-quantum group related
to the Alexander-Conway polynomial
\footnote{Talk given at the XXVIII Karpacz Winter School, Karpacz,
Poland, February 1992}\\
\vspace{0.45in}
\ns\sc Shahn Majid
\footnote{SERC (UK) Research Fellow and Drapers Fellow of Pembroke
College, Cambridge}\\
\ssize\em
Department of Applied Mathematics and
Theoretical Physics\\
University of Cambridge, Silver Street\\
Cambridge CB3 9EW, UK\\
\vspace{0.2in}
\ns\sc
M.J. Rodr\'\i guez-Plaza
\footnote{CSIC (Spain) Research Fellow}\\
\ssize\em
The Niels Bohr Institute\\
Blegdamsvej 17\\
DK-2100 Copenhagen \O, Denmark\\
\vspace{0.2in}
\end{center}

\vglue 4 true cm

{\leftskip=1.5 true cm \rightskip=1.5 true cm
\openup 1\jot
{\centerline{\bf Abstract}}
\vspace{0.2in}
\noindent
We describe the quasitriangular structure (universal $R$-matrix)
on the non-standard quantum group $\setu$ associated to the
Alexander-Conway matrix solution of the Yang-Baxter equation.
We show that this Hopf algebra is connected with the
super-Hopf algebra $\super$ by a general process of superization.
\\
\par}


\openup 2\jot
\section{Introduction}

It is often associated the term quantum group with the Quantum
Algebras $U_q{\g}$ of Drinfeld and Jimbo but the theory of
quasitriangular Hopf algebras (quantum groups)
covers other non-standard examples. New quantum groups can be generated
with the quantum double construction of Drinfeld \cite{Dri}
or the FRT approach \cite{FRTlie}, though it shall not believe
that all the known quantum groups follow necessarily one of these two
methods.
Since only the FRT construction will be our concern here
we refer to it briefly. To any standard matrix solution $R$
of the Quantum Yang-Baxter Equation (QYBE), the FRT approach associates
two bialgebras $A(R)$ of `function algebra' type and $U(R)$ of
`enveloping algebra' type that under certain conditions can lead to a Hopf
algebra and a quantum group. This was done in the pioneering work of
\cite{FRTlie} providing in this way
a matrix description of the standard $U_q{\g}$.
However, the construction is far more general as it was shown in
\cite{Maqua}. In this reference it is established
that for quite general solutions $R$ it is possible to
make $A(R)$ into a Hopf algebra and that the resulting quantum group of
enveloping algebra type is necessarily quasitriangular, i.e. possessing a
`universal $R$-matrix'. This indicates that the FRT construction can be
extended to a large class of new quantum groups since many
`non-standard' solutions $R$ are known.

Here we develop an example of this type. We study the Hopf algebra
$\,U=\setu$ associated to the non-standard solution
\begin{eqnarray}
R=\l(\begin{array}{cccc}
q    &    0    &    0        &    0       \\
0    &    1    & q-q^{-1}    &    0       \\
0    &    0    &    1        &    0       \\
0    &    0    &    0        &    -q^{-1}
\label{AC}
\end{array}\r)
\end{eqnarray}
of the matrix QYBE
$R_{12}R_{13}R_{23}=R_{23}R_{13}R_{12}$. This solution gives rise to the
Alexander-Conway (AC) knot polynomial \cite{CLScon} and was first found
in the context of statistical mechanics ($q$-state vertex model) in
\cite{PerSchnew}. Although the relevant new Hopf algebra structure
$\setu$ associated to this Alexander-Conway solution was computed in
\cite{JGWnew} the questions of universal $R$-matrix and superization
were left open. These two problems are solved here.

\section{The quasitriangular Hopf algebra $\setu$}

The algebra $\setu$ is introduced by the following
\newtheorem{defn}{Definition}
\begin{defn}
$\setu$ is the Hopf algebra generated by
the operators $\{\ho,\htw,\xpm\}$, and relations
\begin{eqnarray}
&&\l[\ho,\htw\r]=0,\nn\\
&&\l[\ho,\xpm\r]=\pm 2\xpm,\qquad \l[\htw,\xpm\r]=\mp 2\xpm,
\label{alg}\\
&&\l[\xp,\xm\r]={\ko\kt-\koi\kti\over \dif}, \qquad\l(\xpm\r)^2=0,\nn
\end{eqnarray}
where the operators $\ko,\kt$ are defined in terms of $\ho,\htw$ as
\begin{equation}
\ko=q^{\ho/2},\qquad \kt=\fact q^{\htw/2}
\label{def}
\end{equation}
and $q$ is an arbitrary parameter. The comultiplication relations are
given by
\begin{eqnarray}
&&\tri \hi=\hi\oti\un +\un\oti\hi, \,\,\, i=1,2 \nn\\
          &&\tri \xp=\xp\oti\ko+\kti\oti\xp,\label{coalg}\\
          &&\tri \xm=\xm\oti\kt+\koi\oti\xm\nn,
\end{eqnarray}
and the antipode $S$ and counit $\veps$ as follows
\begin{eqnarray*}
&&S\l(\hi\r)=-\hi,\qquad S\l(\xp\r)=-q\koi\kt\xp,\qquad
            S\l(\xm\r)=q\ko\kti\xm,\\
&&\veps\l(\hi\r)=\veps\l(\xpm\r)=0.
\end{eqnarray*}
\end{defn}

The algebra mentioned above has been constructed using the FRT approach
to obtain bialgebras, or Hopf algebras if possible, from matrix
solutions of the QYBE \cite{FRTlie}. More specifically the set of relations
(\ref{alg}) are
$R_{21}L^\pm_1L^\pm_2=L^\pm_2L^\pm_1R_{21}$ and
$R_{21}L^+_1L^-_2=L^-_2L^+_1R_{21}$, where
$L^\pm_1$ and $L^\pm_2$ denote the operators
$L^\pm_1=L^\pm\oti\un$, $L^\pm_2=\un\oti L^\pm$ and
$L^+,L^-$ are the lower and upper triangular matrices with ansatz
\begin{eqnarray*}
L^+=\l(\begin{array}{cc}
\ko                     &    0               \\
\l(\dif\r)\xp           &    \kti
\end{array}\r),
\qquad\qquad
L^-=\l(\begin{array}{cc}
\koi               &      -\l(\dif\r)\xm     \\
0                       &   \kt
\end{array}\r).
\end{eqnarray*}
The matrix $R_{21}$ is defined as $PRP$, with $P$ is the permutation operator
$P\l(a\oti b\r)=b\oti a$ that in this case presents the form
\[
P=\l(\begin{array}{cccc}
1    &    0    &    0        &    0       \\
0    &    0    &    1        &    0       \\
0    &    1    &    0        &    0       \\
0    &    0    &    0        &    1
\end{array}\r)
\]
and $R$ is the AC solution shown in (\ref{AC}). The coalgebra structure
(\ref{coalg}) comes from the relation
$\tri L^\pm_{ij}=\sum_k L^\pm_{ik}\oti L^\pm_{kj}$, where $L^\pm_{ij}$
are the matrix elements of $L^\pm$. The formula $\l(\xpm\r)^2=0$
exhibited in (\ref{alg}) are highly suggestive of a superalgebra
with $\xpm$ odd and $\ko,\kt$ even operators. However, $\setu$ is not a
super-quantum group. The reason is simple. To extend the
comultiplication $\tri$ to products of generators we must use the
bosonic manipulation $(a\oti b)(c\oti d)=ac\oti bd$
with respect to which the map
$\tri: U\longrightarrow U\oti U$ is an algebra homomorphism. If $U$ were
a super-quantum group this relation should be substituted by
$(a\oti b)(c\oti d)=(-1)^{{\rm deg}(b){\rm deg}(c)} ac\oti bd$, which is
inconsistent with the relations in the algebra $U$. In other words,
$\setu$ reminds us of a super-quantum group but it is an ordinary
bosonic one. We address this problem the next section.

Let us present now the quasitriangular structure for this Hopf algebra
$\setu$. We recall that any Hopf algebra $U$ is called {\em quasitriangular}
if there is an invertible element $\CR$ in $U\oti U$ that obey the
axioms \cite{Dri}
\begin{equation}
\tri '\l(a\r)=\CR\tri\l(a\r)\CRi
\label{cobound}
\end{equation}
for all $a$ in $U$ and
\begin{equation}
\l(\tri\oti{\rm id}\r)\CR=\CR_{13}\CR_{23},\qquad
\l({\rm id}\oti\tri\r)\CR=\CR_{13}\CR_{12}.
\label{drinf}
\end{equation}
Here $\tri '$ is defined as
$\tau\circ\tri$ where $\tau(x\oti y)\mapsto y\oti x$ and the object
$\CR$ is known as the `{\em universal $R$-matrix}'.
Equations (\ref{drinf}) are evaluated in $U^{\oti 3}$ and if we write
$\CR$ as the formal sum $\CR=\sum_i a_i\oti b_i$, the operators
$\CR_{12}, \CR_{13}$ and $\CR_{23}$ are defined by the expressions
$\CR_{12}=\sum_i a_i\oti b_i\oti\un$,
$\CR_{13}=\sum_i a_i\oti \un\oti b_i$ and
$\CR_{23}=\sum_i\un\oti a_i\oti b_i$.
The physical importance of $\CR$ is that it satisfies
the quantum Yang-Baxter equation
\[
\CR_{12}\CR_{13}\CR_{23}=\CR_{23}\CR_{13}\CR_{12}
\]
abstractly in the algebra $U$. Now we find such a quasitriangular
structure
\newtheorem{thm}{Theorem}
\begin{thm}

The Hopf algebra $\setu$ is quasitriangular with
universal $R$-matrix
\begin{equation}
\CR=e^{-i{\pi\over 4}\htw\oti\htw}\,
q^{{1\over 4}\l(\ho\oti\ho-\htw\oti\htw\r)}\l(\un\oti\un+\l(1-q^2\r) E\oti
F\r),
\label{R}
\end{equation}
where $E$ and $F$ denote the operators $\kt\xp$ and $\kti\xm$,
respectively.
\end{thm}

\newtheorem{proof}{Proof}
\begin{proof}
It is sufficient to show that this $\CR$ verifies Drinfeld's axioms
(\ref{cobound}) and (\ref{drinf}).
$\Box$
\end{proof}

This quasitriangular structure is crucial for numerous applications. In
particular, the quantity $u=\sum_i S(b_i)a_i$ which implements the square
of the antipode in the form $uau^{-1}=S^2(a)$ for all elements $a$ of
the quantum group is of special interest in the general theory. In this
case $u$ in $\setu$ is given by
\[u=e^{i{\pi\over 4}\htw^2}q^{-\cuar\l(\ho^2-\htw^2\r)}
\l(\un+\l(1-q^2\r)\koi\kti FE\r).
\]
Another motivation to evaluate the elements $R$ and $u$ is that they are
useful in providing link invariants.

We finish this section referring to the
{\em canonical representation} of $\setu$ defined by
\begin{eqnarray}
&&\rho(\ho)=\l(\begin{array}{cc}
 2  &   0  \\
 0  &   0
\end{array}\r),\qquad
\rho(\htw)=\l(\begin{array}{cc}
0   &   0   \\
0   &   2
\end{array}\r),\nn\\
&&\rho(\xp)=\l(\begin{array}{cc}
0   &   1   \\
0   &   0
\end{array}\r),\qquad
\rho(\xm)=\l(\begin{array}{cc}
0   &   0   \\
1   &   0
\end{array}\r).
\label{canonical}
\end{eqnarray}
In this canonical representation $\rho$ we recover
$\rho\oti\rho\l(\CR\r)=R$, the AC solution (\ref{AC}) of the
introduction. This result is to be expected from the general theory
of matrix quantum groups \cite[Sec. 3]{Maqua} and also serves as a
check on our universal $R$-matrix (\ref{R}).

\section{Superization of $\setu$}

We have already pointed that the relations $\l(\xpm\r)^2=0$ in
(\ref{alg}) are surely
indicative of some kind of super-Hopf algebra structure.
Yet $\setu$ is an ordinary Hopf algebra and therefore not a
super one at all. We give now some insight into this puzzle
by means of the transmutation theory of \cite{Matra}. This reference
contains theorems which show how, under suitable circumstances,
we can transform Hopf algebras into super-Hopf algebras and vice-versa.
The purpose of this section is to
prove, using one of these theorems, that the superization of
a quotient of $\setu$ coincides with the super-quantum group $\super$.
We begin by introducing the concept of superization of an
ordinary Hopf algebra.
If $H$ is any Hopf algebra containing a group-like element $g$ such that
$g^2=\un$, we can define its {\em superization} as the super-Hopf algebra
$\und H$ with the following properties.
As an algebra $\und H$ coincides with $H$ and the counit also
coincides. As far as the comultiplication, antipode and
quasitriangular structure (if any) of $H$ are concerned,
all are modified to the super ones given by
\[
\und \tri h=\sum_k x_kg^{-{\rm deg} \l(y_k\r)}\oti y_k,\qquad
{\und S}\l(h\r)=g^{{\rm deg}\l(h\r)} S\l(h\r)
\]
and
\[
\und \CR=\CR_g^{-1}\sum_i a_ig^{-{\rm deg} \l(b_i\r)}\oti
 b_i.
\]
Here $h$ denotes an arbitrary element of $H$ with comultiplication
$\tri h=\sum_k x_k\oti y_k$,
$\CR=\sum_i a_i\oti b_i$ is the universal $R$-matrix of $H$
and the element $\CR_g$ is $\CR_g=\CR_g^{-1}=\un\oti\un-2p\oti p$,
with $p=\l(\un-g\r)/2$. When $h$ is
considered as an element of $\und H$ its grading ${\rm deg}\l(h\r)$ is
given by the action of $g$ on $h$ in the adjoint representation, that is
to say by $ghg^{-1}={\rm deg}\l(h\r)\cdot h$ on homogeneous elements.

In order to apply the above superization theorem to $\setu$ we first note
that in this algebra the role of
$g$ can be played by the operator $e^{{i\pi\htw}/2}$ introduced in
(\ref{def}).
This element $g=e^{{i\pi\htw}/2}$ has the property that $g^2$ is
central and group-like. Hence it is natural to impose the relation
$g^2=\un$ in the abstract algebra. Moreover, we know that
$\rho(g^2)=\un$ in the representation (\ref{canonical}),
so this further relation is consistent with the canonical
representation. Note that $g$ itself commutes with $\ko,\kt$
and anticommutes with $X^\pm$.
We have

\newtheorem{prop}{Proposition}
\begin{prop}
Let $g=e^{{i\pi\htw}/2}$. The superization of the Hopf
algebra $\setu/{g^2-\un}$ is the super-Hopf algebra $\super/e^{2\pi iN}-\un$.
Here $\super$ is defined by generators $C,N$ even and
$\eta$, $\etap$ odd with relations
\begin{eqnarray}
&&[N,\eta]=-\eta ,\qquad [N,\etap]=\etap,\nn\\
&&\l\{\eta,\etap\r\}={q^C-q^{-C}\over \dif},\label{superalg}\\
&&\eta^2=0,\qquad \l(\etap\r)^2=0\nn,
\end{eqnarray}
and the operator $C$ central. The supercomultiplication is given by
\begin{eqnarray}
&&\und \tri C=C\oti \un+\un\oti C,\qquad
\und \tri N=N\oti\un+\un\oti N,\nn\\
&&\und \tri \eta=\eta\oti q^{C-N}+q^{-N}\oti\eta,\qquad
\und \tri \etap=\etap\oti q^N+q^{-C+N}\oti\etap.
\label{supercoalg}
\end{eqnarray}
This $\super$ has a super-quasitriangular structure given by the
expression
\begin{equation}
\und\CR=q^{-\l(C\oti N+N\oti C\r)}
\l(\un\oti\un+\l(1-q^2\r)q^N\eta\oti q^{-N}\etap\r).
\label{superR}
\end{equation}
\end{prop}

\begin{proof}
The proof of this proposition follows from a straightforward
application of the above superization construction applied to the
quotient. According to this the outcoming super-Hopf algebra
is generated by $\ho,\,\htw$ even and
$\xpm$ odd operators with relations (\ref{alg}) unchanged,
supercomultiplication given by
\begin{eqnarray*}
&&\und \tri H_i=H_i\oti\un+\un\oti H_i,\qquad i=1,2,\\
&&\und \tri \xp=\xp\oti\ko+\kti g\oti\xp,\\
&&\und \tri \xm=\xm\oti\kt+\koi g\oti\xm,
\end{eqnarray*}
and superantipode and supercounit as follows
\begin{eqnarray*}
&&\und S\l(\hi\r)=-\hi,\qquad \und S\l(\xp\r)=-q\koi\kt g\xp,\qquad
            \und S\l(\xm\r)=q\ko\kti g\xm,\\
&&\und \veps\l(\hi\r)=\und\veps\l(\xpm\r)=0.
\end{eqnarray*}
In these expressions the $K_i$ are defined as in (\ref{def}). It is not
difficult to recognize the resulting super-quantum group as
$\super$ if we redefine the generators
$H_1,H_2,X^\pm$ as $C,N,\eta,\etap$ according to
\[ C=\l(\ho+\htw\r)/2,\qquad N=\htw/2,\qquad \eta=\xp,
\qquad \etap=\xm g
\]
(so that $q^C=\ko\kt g$ and $q^N=\kt g$).
Note that a direct consequence
of this definition is that $C$ is central as stated.
Introducing these definitions in the above relations we obtain
(\ref{superalg}), (\ref{supercoalg}) and (\ref{superR})
without difficulty.
Let us stress that the supercomultiplication (\ref{supercoalg})
is an algebra homomorphism consistent with the relations
(\ref{superalg}), provided we use super manipulation.
$\Box$
\end{proof}

The super-quantum group $\setu$ has already been connected
with the Alexander-Conway polynomial in a physical state in
\cite{KauSalfre} so the importance of this proposition lies in the fact
that it solves precisely the suspected connection between
$\setu$ and $\super$.

We conclude this section mentioning that the previous superization
theorem is not the only existing method of turning quantum groups into
super-quantum groups. This $\super$ can also be obtained by the {\em
graded} FRT construction associated to the graded variant of the solution
(\ref{AC}) \cite{MaRo}. This confirms that the
superization of $\setu$ shown here is fully compatible with
existing ideas of transforming solutions of the QYBE into solutions
of the graded QYBE and finding their corresponding quantum groups.


\end{document}